\documentclass[journal]{IEEEtran}
\usepackage{amsfonts}

\usepackage{bm}
\usepackage{amsmath}
\usepackage{makecell}
\usepackage{amssymb}
\usepackage{array}
\usepackage{cases}
\usepackage{algorithm}
\usepackage{algorithmic}
\usepackage{enumerate}
\usepackage{color,xcolor}
\usepackage{graphicx}
\usepackage{multirow,tabularx}
\usepackage{subfigure}
\usepackage{multirow}
\usepackage{hyperref}
\usepackage[compress]{cite}
\usepackage{setspace}

\makeatletter

\newcommand{\Rmnum}[1]{\expandafter\@slowromancap\romannumeral #1@}
\usepackage[vlined, ruled, algo2e]{algorithm2e}
\makeatother
 0

\begin{document}
%\title{Optimal Diversity-Multiplexing Tradeoff \\of MIMO Multi-Way Relay Channel}
\title{DNN-Aided Message Passing Based Block Sparse Bayesian Learning for Joint User Activity Detection and Channel Estimation}

\author{
\IEEEauthorblockN{Zhaoji~Zhang, Ying~Li, \IEEEmembership{Member, IEEE}, Chongwen Huang, Qinghua Guo, \IEEEmembership{Senior Member, IEEE}, Chau Yuen, \IEEEmembership{Senior Member, IEEE}, and Yong Liang Guan, \IEEEmembership{Senior Member, IEEE}}
}

\maketitle
\begin{abstract}
Faced with the massive connection, sporadic transmission, and small-sized data packets in future cellular communication, a grant-free non-orthogonal random access (NORA) system is considered in this paper, which could reduce the access delay and support more devices. In order to address the joint user activity detection (UAD) and channel estimation (CE) problem in the grant-free NORA system, we propose a deep neural network-aided message passing-based block sparse Bayesian learning (DNN-MP-BSBL) algorithm. In this algorithm, the message passing process is transferred from a factor graph to a deep neural network (DNN). Weights are imposed on the messages in the DNN and trained to minimize the estimation error. It is shown that the weights could alleviate the convergence problem of the MP-BSBL algorithm. Simulation results show that the proposed DNN-MP-BSBL algorithm could improve the UAD and CE accuracy with a smaller number of iterations.
\end{abstract}
\begin{IEEEkeywords}
deep neural network, sparse Bayesian learning, grant-free, user activity detection, channel estimation
\end{IEEEkeywords}


% \fi
%\IEEEpeerreviewmaketitle

\section{Introduction}
\IEEEPARstart{P}{roviding} efficient support for the Internet of Things (IoT) is one of the major objectives for the cellular wireless communication. Machine-to-Machine (M2M) communication is anticipated to support billions of Machine Type Communication (MTC) devices. In addition, the MTC devices are sporadically activated with short packets \cite{servicetype}. Therefore, the random access (RA) process for M2M communications in IoT is characterized by massive connection and sporadic transmission, as well as small-sized data packets.

Confronted with the characteristics described above, the conventional orthogonal multiple access (OMA) scheme becomes infeasible due to its low resource efficiency. To facilitate the sharing of uplink resources, different RA schemes were proposed and can be generally categorized into two types: grant-based RA \cite{ACB,seperation,dynamicone} and grant-free RA \cite{CRAN,2015ICC,ICASSP,AMPmassive,SBLAMP,LSAMPSBL,MPBSBL2019,MPBSBL}.
\subsection{Grant-Based Random Access}
In grant-based RA schemes, activated users contend for the RBs by transmitting a preamble sequence to the base station (BS), while a RB is assigned to the activated user, whose preamble sequence is received and accepted by the BS. One problem with the grant-based RA schemes is that the RB is wasted when more than one active device transmit the same preamble sequence. Some solutions were proposed to alleviate the RA congestion by reducing the collision probability, such as the Access Class Barring (ACB) scheme \cite{ACB}, delicate splitting of the RA preamble set \cite{seperation}, and automatic configuration of the RA parameters \cite{dynamicone}. However, the RB wastage cannot be fully avoided by grant-based RA schemes, which results in low resource efficiency. Furthermore, a handshaking process is always to recognize the contention winner, which undermines the uplink transmission efficiency of small data packets.
\subsection{Grant-Free Random Access}
\subsubsection{Compressed sensing-based grant-free RA schemes}
Compressed sensing (CS) algorithms employ pilot sequences to accomplish the user activity detection (UAD) and/or channel estimation (CE) problem. For example, the joint UAD and CE problem was addressed by a modified Bayesian compressed sensing algorithm \cite{CRAN} for the cloud radio access network (C-RAN). In addition, the powerful approximate message passing (AMP) algorithm was employed for the joint UAD and CE problem when the BS is equipped either with a single antenna \cite{2015ICC, ICASSP} or with multiple antennas \cite{AMPmassive}.
\subsubsection{Sparse Bayesian learning-based grant-free RA schemes}
The sparse Bayesian learning (SBL) algorithm considers the prior hyper-parameter of the sparse signal. The Expectation Maximization (EM) method was employed by the AMP-SBL algorithm \cite{SBLAMP} to update the sparse signal and the hyper-parameter. A least square (LS)-based AMP-SBL (LS-AMP-SBL) algorithm \cite{LSAMPSBL} was proposed to recover the sparse signal in three steps. Recently, a message-passing receiver design was proposed for the joint channel estimation and data decoding in uplink grant-free SCMA systems \cite{MPBSBL2019}. In addition, a message passing-based block sparse Bayesian learning (MP-BSBL) algorithm \cite{MPBSBL} was proposed for a grant-free NOMA system.
\begin{figure*}
\centering
\includegraphics[width=0.75\linewidth]{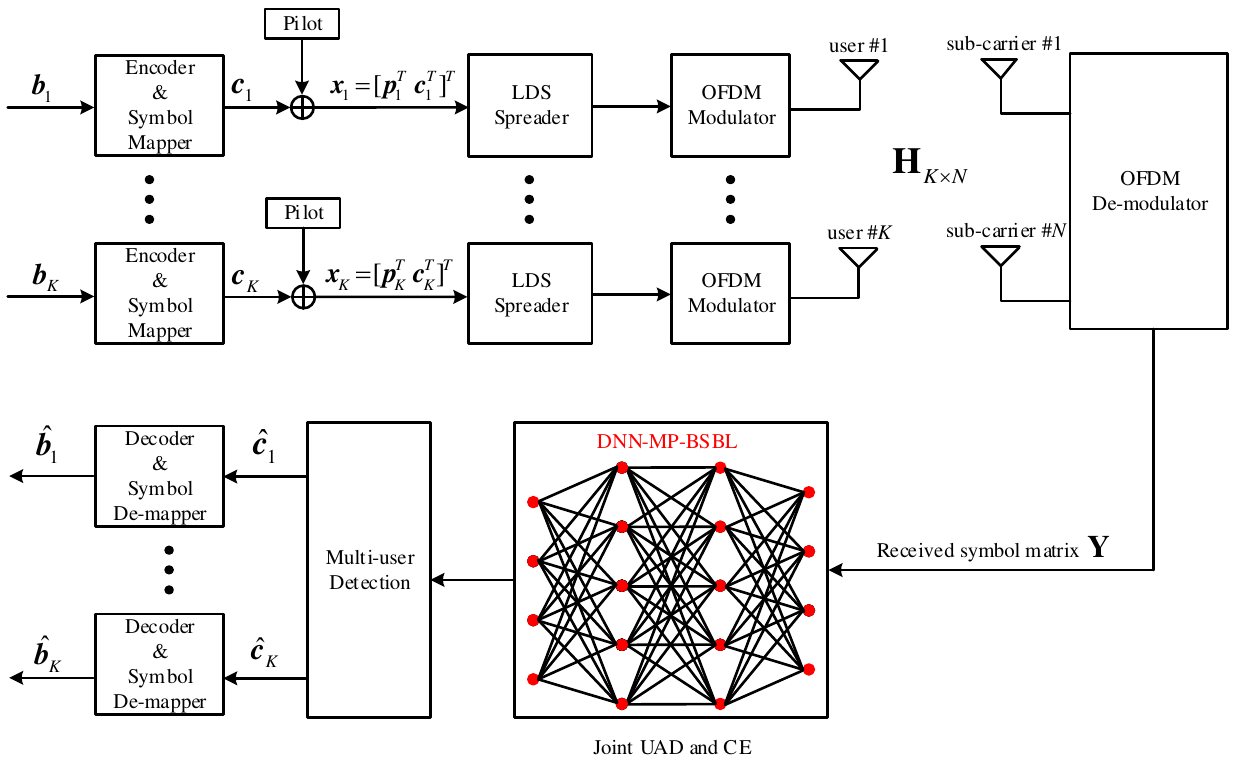}\vspace{-0.3cm}
\caption{LDS-OFDM based grant-free NORA system model. The joint UAD and CE problem is sloved by the proposed DNN-MP-BSBL algorithm, which processes the first $L_t$ received signals of the pilot sequences, while the MUD module is employed to process the remaining $L_c$ received signals to detect the data transmitted from active users.}\vspace{-0.3cm}
\label{SystemModel}
\end{figure*}
\subsection{Contributions}
In this paper, we consider a LDS-OFDM system, where devices perform grant-free RA once they are activated. A deep neural network-aided message passing-based block sparse Bayesian learning (DNN-MP-BSBL) algorithm is proposed to perform joint UAD and CE. The iterative message passing process of the MP-BSBL algorithm \cite{MPBSBL} is transferred from a factor graph to a neural network. Weights are imposed on the messages passing in the neural network and trained to minimize the estimation error.
%It is shown that the trained weights could alleviate the convergence problem of the MP-BSBL algorithm, especially in crowded RA scenarios. As a result, the DNN-MP-BSBL algorithm could improve the UAD and CE accuracy with a smaller number of iterations. In addition, the training of the DNN is conducted offline, i.e., the DNN-MP-BSBL algorithm does not lead to any additional online computational complexity. This guarantees the feasibility of the proposed DNN-MP-BSBL algorithm for crowded NORA systems with real-time implementation and low-latency requirement.

The rest of this paper is organized as follows. The system model and the MP-BSBL algorithm are presented in Section \ref{ModelProblem}. The DNN structure for the DNN-MP-BSBL algorithm is illustrated in Section \ref{DNN_SBSL}, where the weighted message passing is explained in details. Simulation results are given in Section \ref{simulationSec} to verify the UAD and CE accuracy of the proposed DNN-MP-BSBL algorithm. Finally, Section \ref{conclusions} concludes this paper.

\section{Joint UAD and CE by MP-BSBL}\label{ModelProblem}
\subsection{System Model and Problem Formulation}
A LDS-OFDM system is considered in Fig. \ref{SystemModel}. There are $N$ sub-carriers and $K$ users. Each user is activated with probability $P_a$. For active user $k$, its information sequence $\mathbf{b}_k$ is encoded and mapped into a QAM sequence $\mathbf{c}_k\in\mathbb{C}^{L_c\times1}$ with length $L_c$. ZC sequences are adopted as pilot sequences. One unique sequence $\mathbf{p}_k$ with length $L_t$ is allocated for user $k$, and inserted into the transmitted sequence $\mathbf{x}_k$, i.e., $\mathbf{x}_k=[\mathbf{p}_k^T \ \mathbf{c}_k^T]^T$. Therefore, the length $L$ of $\mathbf{x}_k$ is $L=L_t+L_c$. The LDS spreader for user $k$ is characterized by $\mathbf{s}_k$, which is a sparse vector with length $N$ and $d_c$ non-zero elements. The LDS spreaders for all the $K$ users are characterized by a LDS spreading matrix $\mathbf{S}=[\mathbf{s}_1,...,\mathbf{s}_K]$. We consider a regular $\mathbf{S}$, i.e., the column degree $d_c$ and the row degree $d_r$ in $\mathbf{S}$ are constant. Each sub-carrier is shared by $d_r$ potential users, with $d_{r}=(K/N)d_{c}$. When multiple users are active on the same sub-carrier, the RA is conducted in a NOMA manner. After the OFDM de-modulator, the received matrix is
\begin{equation}\label{received}
\mathbf{Y}_{L\times N}=\mathbf{X}_{L\times K}\mathbf{H}_{K\times N}+\mathbf{W}_{N\times L},
\end{equation}
where the $(l,n)$-th entry of $\mathbf{Y}$ represents the $l$-th received symbol on the $n$-th sub-carrier, the $(l,k)$-th entry of $\mathbf{X}$ represents the $l$-th transmitted symbol of the $k$-th user, and $\mathbf{H}=[\alpha_1\mathbf{h}_1,...,\alpha_K\mathbf{h}_K]^T$ is a $K\times N$ row sparse channel matrix, which integrates the activity of the users. The activity indicator $\alpha_k=0$ if user $k$ is inactive. Otherwise, $\alpha_k=1$ and the $k$-th row $\mathbf{h}^T_k$ of $\mathbf{H}_{K\times N}$ represents the channel gain vector on $N$ sub-carriers for user $k$. The entries in the AWGN matrix $\mathbf{W}$ are assumed i.i.d with noise variance $\sigma_{w}^2$. $\mathbf{Y}$ can be decomposed as $\mathbf{Y}_{L\times N}=\left[(\mathbf{Y}^P_{L_t\times N})^T\ (\mathbf{Y}^c_{L_c\times N})^T \right]^T$, where $\mathbf{Y}^P$ and $\mathbf{Y}^c$ represent the received matrices w.r.t. the pilot sequences and the data sequences, respectively. We consider $\mathbf{Y}^P$ to solve the joint UAD and CE problem,
\begin{equation}\label{Pilotreceived}
\mathbf{Y}^P_{L_t\times N}=\mathbf{P}_{L_t\times K}\mathbf{H}_{K\times N}+\mathbf{W}_{N\times L_t},
\end{equation}
where $\mathbf{P}$ is assumed known to the receiver. Then we perform vectorization on the transpose of $\mathbf{Y}^P$ as in \cite{MPBSBL},
\begin{equation}\label{Vectorization}
\begin{split}
\mathbf{y}&=vec([{\mathbf{Y}^P}]^T)=(\mathbf{P}\otimes\mathbf{I}_N)vec(\mathbf{H}^T)+\mathbf{w}\\
&=\!\!\underbrace{\left[ {\begin{array}{*{20}{c}}
p_{1,1}\mathbf{I}_N&p_{1,2}\mathbf{I}_N&\cdots&p_{1,K}\mathbf{I}_N\\
p_{2,1}\mathbf{I}_N&p_{2,2}\mathbf{I}_N&\cdots&p_{2,K}\mathbf{I}_N\\
\vdots&\vdots&\ddots&\vdots\\
p_{L_t,1}\mathbf{I}_N&p_{L_t,2}\mathbf{I}_N&\cdots&p_{L_t,K}\mathbf{I}_N
\end{array}} \right]}_{\mathbf{P}_s}\!
\underbrace{\left[ {\begin{array}{*{20}{c}}
\alpha_1\mathbf{h}_1\\
\alpha_2\mathbf{h}_2\\
\vdots\\
\alpha_K\mathbf{h}_K
\end{array}} \right]}_{\mathbf{h}_s}\!\!+\mathbf{w}\\
&\overset{(a)}{=}\overline{\mathbf{P}}
\underbrace{\left[ {\begin{array}{*{20}{c}}
\alpha_1\mathbf{\overline{h}}_1\\
\alpha_2\mathbf{\overline{h}}_2\\
\vdots\\
\alpha_K\mathbf{\overline{h}}_K
\end{array}} \right]}_{\mathbf{\overline{h}}}+\mathbf{w},
\end{split}
\end{equation}
where $\mathbf{P}_s=\mathbf{P}\otimes\mathbf{I}_N$, $\mathbf{h}_s=vec(\mathbf{H}^T)$, and $\mathbf{h}_k$ is the channel gain vector of user $k$ on $N$ sub-carriers. According to the LDS spreading matrix $\mathbf{S}$, the transmitted symbols of each user are only spread onto $d_c$ sub-carriers. Therefore, $N-d_c$ elements in $\mathbf{h}_k$ are zero. We further simplify $\mathbf{h}_s$ by eliminating the zeros. Accordingly, the columns in $\mathbf{P}_s$ corresponding to the zeros in $\mathbf{h}_s$ are also removed. Finally, we obtain the simplified version of (\ref{Pilotreceived}) in equation ($a$) of (\ref{Vectorization}). According to (\ref{Vectorization}), the joint UAD and CE problem is equivalent to recovering $\overline{\mathbf{h}}$.
%\begin{remark}
%To enable this joint UAD and CE task, each active user transmits its unique pilot sequence before its data transmission. A common practice is to employ the Zadoff-Chu (ZC) sequences \cite{ZC} as pilot sequences, which is also considered in this work. One ZC sequence with length $L_t$ is generated as follows,
%\begin{equation}\label{ZC}
%\mathbf{z}_{u}(n)=\exp\{-i\pi un(n+1)/L_t\},\ \ 0\leq n<L_t,
%\end{equation}
%where $i^2=-1$, the sequence length $L_t$ is a prime number, and $u \in [1,...,L_t-1]$ is the root of $\mathbf{z}_{u}(n)$. When the root $u$ is fixed, we can generate $N_u=\lfloor L_t/N_{SF}\rfloor$ different pilot sequences by cyclically shifting one existing sequence by $N_{SF}$ elements. Therefore, when $N_{SF}$ is set to 1, $L_t(L_t-1)$ different pilot sequences can be generated for the users.
%\end{remark}
\subsection{MP-BSBL Algorithm \cite{MPBSBL}}
The recovery of $\mathbf{\overline{h}}$ can be addressed by the MP-BSBL algorithm \cite{MPBSBL}.
\begin{figure}
\centering
\includegraphics[width=0.87\linewidth]{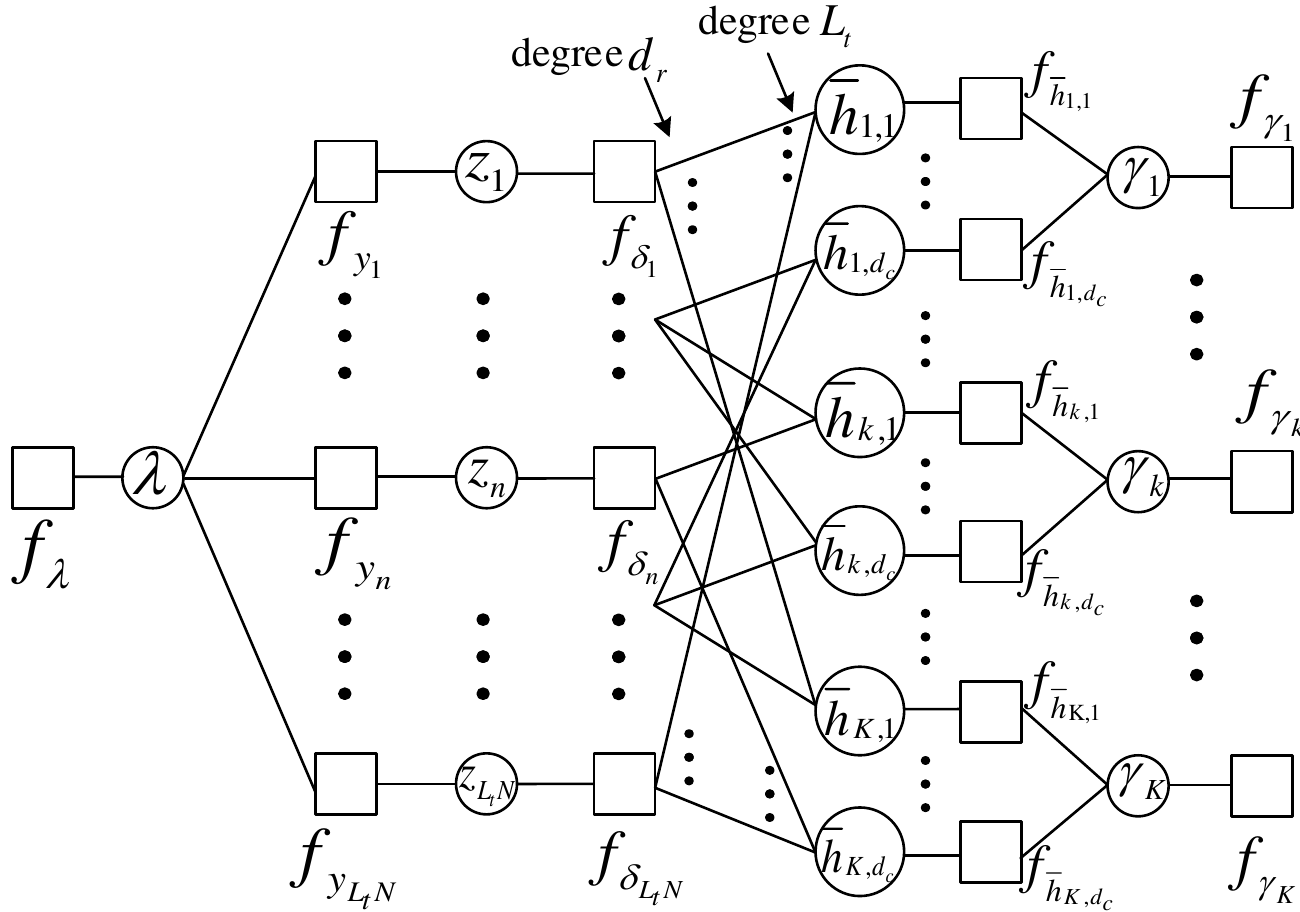}\vspace{-0.3cm}
\caption{Factor graph for the message passing in the MP-BSBL algorithm \cite{MPBSBL}.}\vspace{-0.2cm}
\label{FactorGraph}
\end{figure}
For user $k$, the distribution of $\mathbf{\overline{h}}_k$ is assumed conditioned on a hyper-parameter ${\gamma}_k$, i.e., ${\mathbf{\overline{h}}_k}({\gamma}_k)\sim \mathcal{CN}(\mathbf{0},{\gamma}_k^{-1}\mathbf{I}_{d_c})$. The hyper-parameter ${\gamma}_k$ is assumed to follow a Gamma distribution. The noise precision $\lambda = 1/{\sigma_w^2}$ is unknown at the receiver but assumed with a priori probability $p(\lambda)$. With these assumptions above, the joint a posterior probability is factorized as follows
\begin{equation}\label{Factorize}
\begin{split}
p(\mathbf{\overline{h}},\gamma,\lambda|\mathbf{y}) &\propto p(\mathbf{y}|\mathbf{\overline{h}},\lambda)p(\mathbf{\overline{h}}|\mathbf{\gamma})p(\lambda)p(\gamma)\\
&=p(\lambda)\prod_{n=1}^{L_tN}p\left(y_n|\mathbf{\overline{h}},\lambda\right)\prod_{k=1}^{K}\prod_{d=1}^{d_c}p(\overline{h}_{k,d}|\gamma_k)p(\gamma_k),
\end{split}
\end{equation}
where $p(\lambda) \propto 1/\lambda$, $p(y_n|\mathbf{\overline{h}},\lambda)=\mathcal{CN}(y_n;\mathbf{\overline{p}^T_n}\mathbf{\overline{h}},\lambda^{-1})$, $\mathbf{\overline{p}^T_n}$ is the $n$-th row of matrix $\overline{\mathbf{P}}$ in (\ref{Vectorization}),  $p(\overline{h}_{k,d}|\gamma_k)=\mathcal{CN}(\overline{h}_{k,d};0,\gamma_k^{-1})$, $p(\gamma_k)=Ga(\gamma_k;a_k,b_k)$, and $p(\gamma)=\prod_{k=1}^{d_c}p(\gamma_k)$, $\mathcal{CN}(x;\mu,\sigma^2)$ represents the complex Gaussian distribution probability density function (pdf) of $x$ with mean $\mu$ and variance $\sigma^2$, while $Ga(x;a,b)$ represents the Gamma distribution pdf of $x$ with parameters $a$ and $b$. The parameters $a$ and $b$ are usually assumed in the order of $10^{-4}$.

A factor graph is established for the MP-BSBL algorithm in Fig. \ref{FactorGraph}, where $f_{\overline{h}_{k,d}}(\overline{h}_{k,d},\gamma_k)$, $f_{\gamma_k}(\gamma_k)$ and $f_{y_n}(\mathbf{\overline{h}},\lambda)$ denote $p(\lambda)$, $p(\overline{h}_{k,d}|\gamma_k)$, $p(\gamma_k)$, and $p({y_n}|\mathbf{\overline{h}},\lambda)$. The extra variable $z_n=\mathbf{\overline{p}^T_n}\mathbf{\overline{h}}$ is introduced and the constraint $\delta(z_n-\mathbf{\overline{p}^T_n}\mathbf{\overline{h}})$ is represented by $f_{\delta_n}$. Then, $f_{y_n}$ is a function of $z_n$ and $\lambda$, i.e., $f_{y_n}(z_n,\lambda)=\mathcal{CN}(y_n;z_n,\lambda^{-1})$. The MP-BSBL algorithm performed on the factor graph in Fig. \ref{FactorGraph} is briefed as follows

Denote $l$ as the iteration index and $Q_{k,d}$ as the product of all the incoming messages from  $\delta_{n'}$ to $\overline{h}_{k,d}$. Then, the variance $v_{Q_{k,d}}$ and mean $m_{Q_{k,d}}$ of $\overline{h}_{k,d}$ are,
\begin{equation}\label{vQkd}
v^{l}_{Q_{k,d}}\approx\left( \sum_{n'\in\mathcal{N}\left(\overline{h}_{k,d}\right)}\frac{\left|\overline{P}_{n',kd}\right|^2}{\left(\hat{\lambda}^{l-1}\right)^{-1}+v^{l-1}_{\delta_{n'} \to z_{n'}}} \right)^{-1}
\end{equation}
\begin{equation}\label{mQkd}
m^{l}_{Q_{k,d}}\approx v^{l}_{Q_{k,d}} \sum_{n'\in\mathcal{N}\left(\overline{h}_{k,d}\right)}\frac{\overline{P}_{n',kd}^{H}\left(y_{n'}-m^{l-1}_{\delta_{n'} \to z_{n'}}\right)}{\left(\hat{\lambda}^{l-1}\right)^{-1}+v^{l-1}_{\delta_{n'} \to z_{n'}}}+m_{\overline{h}_{k,d}}^{l-1}
\end{equation}
The variance $v^{l}_{\overline{h}_{k,d}}$ and mean $m^{l}_{\overline{h}_{k,d}}$ of $\overline{h}_{k,d}$ are updated as
\begin{equation}\label{vmhkd}
\begin{split}
v^{l}_{\overline{h}_{k,d}}&=\frac{1}{\left(v_{Q_{k,d}}^{l}\right)^{-1}+\hat{\gamma}_k^{l-1}}\\
m^{l}_{\overline{h}_{k,d}}&=\frac{m^{l}_{Q_{k,d}}}{1+v^{l}_{Q_{k,d}}\hat{\gamma}_k^{l-1}}
\end{split}
\end{equation}
The variance $v_{\delta_n \to z_n}$ and mean $m_{\delta_n \to z_n}$ from $\delta_n$ to $z_n$ are
\begin{equation}\label{delta2z}
\begin{split}
v^{l}_{\delta_n \to z_n}\! &\approx \!\sum_{ \left\{ i,j\right\} \in \mathcal{N}\left(f_{\delta_n}\right)  }\left| \overline{P}_{n,ij}\right|^2v^{l}_{\overline{h}_{i,j}}\\
m^{l}_{\delta_n \to z_n}\! &\approx \!\sum_{ \left\{ i,j\right\} \in \mathcal{N}\left(f_{\delta_n}\right)  }\overline{P}_{n,ij}m^{l}_{\overline{h}_{i,j}}\!-\!\frac{v^{l}_{\delta_n \to z_n}\left( y_n\!-\!m_{\delta_n \to z_n}^{l-1}\right)}{\left(\hat{\lambda}^{l-1}\right)^{-1}\!+\!v_{\delta_n \to z_n}^{l-1}}
\end{split}
\end{equation}
Then $\hat{\gamma}^{l}_k$ is updated by the MF message passing,
\begin{equation}\label{gamma}
\hat{\gamma}^{l}_k=\frac{a_k+d_c+1}{b_k+\sum\limits_{d=1}^{d_c}\left( \left|m^{l}_{\overline{h}_{k,d}}\right|^2+v^{l}_{\overline{h}_{k,d}}\right)}
\end{equation}
The variance $v^{l}_{z_n}$ and mean $m^{l}_{z_n}$ of $z_n$ are
\begin{equation}\label{zn}
\begin{split}
v^{l}_{z_n}&=\left(\hat{\lambda}^{l-1}+\left(v^{l}_{\delta_n \to z_n}\right)^{-1}\right)^{-1}\\
m^{l}_{z_n}&=v^{l}_{z_n}\left(y_n\hat{\lambda}^{l-1}+\frac{m^{l}_{\delta_n \to z_n}}{v^{l}_{\delta_n \to z_n}}\right)
\end{split}
\end{equation}
Then, $\hat{\lambda}^{l}$ is updated by MF message passing,
\begin{equation}\label{lambda}
\hat{\lambda}^{l}=\frac{L_tN}{\sum\limits_{n=1}^{L_tN}\left[ \left(m^{l}_{z_n}-y_n\right)^2+v^{l}_{z_n}\right]}
\end{equation}
If $\left({\hat{\gamma}^l_k}\right)^{-1}<\gamma_{th}$, user $k$ is detected as inactive. Otherwise, $\{m^l_{\overline{h}_{k,d}}, d=1,\ldots,d_c\}$ is the estimated channel gain.
%In (\ref{vQkd}) and (\ref{mQkd}), $Q_{k,d}$ represents the product of all the incoming messages from the neighboring nodes $\delta_{n'}$ of $\overline{h}_{k,d}$ while $v_{Q_{k,d}}$ and $m_{Q_{k,d}}$ are the variance and mean of $\overline{h}_{k,d}$ derived from $Q_{k,d}$. In (\ref{vmhkd}), $v_{\overline{h}_{k,d}}$ and $m_{\overline{h}_{k,d}}$ are the variance and mean of $\overline{h}_{k,d}$ derived from $Q_{k,d}$ and the message passed from $f_{\overline{h}_{k,d}}$ to $\overline{h}_{k,d}$. In (\ref{delta2z}), $v_{\delta_n \to z_n}$ and $m_{\delta_n \to z_n}$ are the variance and mean of $z_n$ derived from the messages passed from $\delta_n$ to $z_n$. In (\ref{zn}), $v_{z_n}$ and $m_{z_n}$ are the variance and mean of $z_n$ derived from the messages passed from $f_{y_n}$ and $\delta_n$ to $z_n$.
\section{DNN-Aided MP-BSBL Algorithm}\label{DNN_SBSL}
\begin{figure*}
\centering
\includegraphics[width=0.82\linewidth]{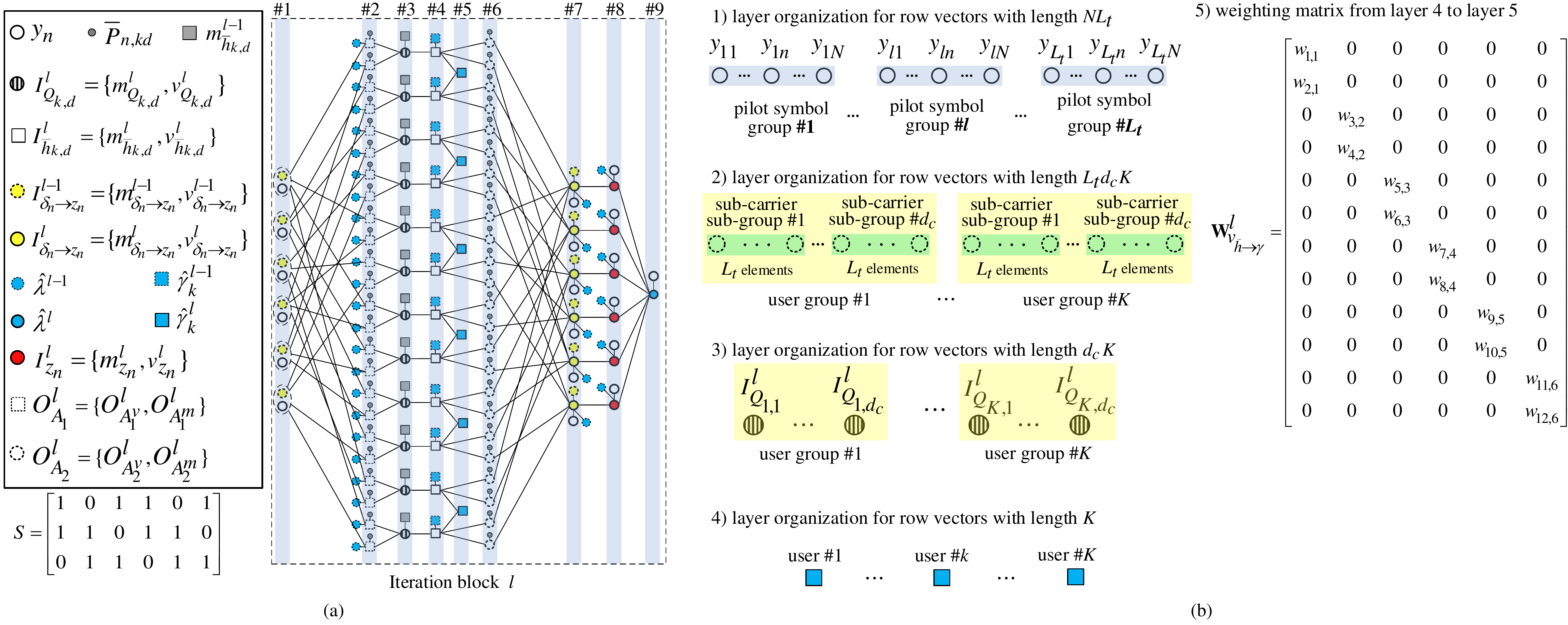}\vspace{-0.35cm}
\caption{DNN for the weighted message passing in DNN-MP-BSBL algorithm with ($N=3, K=6, L_t=2, d_c=2$) and layer organization.}
\label{DNN-MP-BSBL}\vspace{-0.3cm}
\end{figure*}
\begin{table}
\renewcommand\arraystretch{1.2}
\caption{Input and Output for Each Layer of the DNN.}\vspace{-0.2cm}
\centering
\begin{tabular}{cccc}
  \Xhline{1.2pt}
\!\!\!\!Index&\!\!\!\!\!\!\!\!\!\!\!\!\!\!Layer Input&\!\!Layer Output&\!\!\!\!Length\!\!\\
  \hline
\!\!\!\!1&\!\!\!\!\!\!\!\!\!\!\!\!\!\!None&\!\!\!\!$\mathbf{y},\mathbf{I}^{l-1}_{\delta \!\to\! z}\!=\!\{\mathbf{v}^{l-1}_{\delta \!\to\! z},\mathbf{m}^{l-1}_{\delta \!\to\! z}\}$&\!\!\!\!$NL_t$\!\!\\
\!\!\!\!2&\!\!\!\!\!\!\!\!\!\!$\mathbf{y},\mathbf{I}^{l-1}_{\delta \!\to\! z},\hat{\lambda}^{l-1},\mathbf{P}$&\!\!\!\!$\mathbf{O}^l_{A_1}\!=\!\{\mathbf{O}^l_{A^v_1},\mathbf{O}^l_{A^m_1}\}$&\!\!\!\!$L_td_cK$\!\!\\
\!\!\!\!3&\!\!\!\!\!\!\!\!\!\!$\mathbf{m}^{l-1}_{\overline{h}},\mathbf{O}^l_{A_1}$&\!\!\!\!$\mathbf{I}^l_Q\!=\!\{\mathbf{v}^l_Q,\mathbf{m}^l_Q\}$&\!\!\!\!$d_cK$\!\!\\
\!\!\!\!4&\!\!\!\!\!\!\!\!\!\!$\mathbf{I}^l_Q,\hat{\gamma}^{l-1}$&\!\!\!\!$\mathbf{I}^l_{\overline{h}}\!=\!\{\mathbf{v}^l_{\overline{h}},\mathbf{m}^l_{\overline{h}}\}$&\!\!\!\!$d_cK$\!\!\\
\!\!\!\!5&\!\!\!\!\!\!\!\!\!\!$\mathbf{I}^l_{\overline{h}}$&\!\!\!\!$\hat{\mathbf{\gamma}}^l$&\!\!\!\!$K$\!\!\\
\!\!\!\!6&\!\!\!\!\!\!\!\!\!\!$\mathbf{P},\mathbf{I}^l_{\overline{h}}$&\!\!\!\!$\mathbf{O}^l_{A_2}\!=\!\{\mathbf{O}^l_{A^v_2},\mathbf{O}^l_{A^m_2}\}$&\!\!\!\!$L_td_cK$\!\!\\
\!\!\!\!7&\!\!\!\!\!\!\!\!\!\!$\mathbf{y},\mathbf{I}^{l-1}_{\delta \!\to\! z},\hat{\lambda}^{l-1},\mathbf{O}^l_{A_2}$&\!\!\!\!$\mathbf{I}^l_{\delta \!\to\! z}\!=\!\{\mathbf{v}^l_{\delta \!\to\! z},\mathbf{m}^l_{\delta \!\to\! z}\}$&\!\!\!\!$NL_t$\!\!\\
\!\!\!\!8&\!\!\!\!\!\!\!\!\!\!$\mathbf{y},\hat{\lambda}^{l-1},\mathbf{I}^l_{\delta \!\to\! z}$&\!\!\!\!$\mathbf{I}^l_{z}\!=\!\{\mathbf{v}^l_{z},\mathbf{m}^l_{z}\}$&\!\!\!\!$NL_t$\!\!\\
\!\!\!\!9&\!\!\!\!\!\!\!\!\!\!$\mathbf{I}^l_{z},\mathbf{y}$&\!\!\!\!$\hat{\lambda}^l$&\!\!\!\!$1$\!\!\\
  \Xhline{1.2pt}
\end{tabular}
\label{LayerMeaning}
\end{table}
The factor graph in Fig. \ref{FactorGraph} is densely connected, which results in the correlation problem of the Gaussian messages \cite{GMPID,GMPCW,HCWTVT}. To address this problem, we propose a DNN-MP-BSBL algorithm to imposes weights on the Gaussian message update and the MF message update. To facilitate the training, the message passing is transferred from the factor graph to a DNN in Fig. \ref{DNN-MP-BSBL}(a). Each iteration of the MP-BSBL algorithm is now represented by one iteration block. Within each iteration block, one layer represents one particular message. Two auxiliary layers $A_1$ and $A_2$ are also added for illustration clarity. Therefore, as listed in Table \ref{LayerMeaning}, there are 9 layers in each iteration block. The layer organization is shown in Fig. \ref{DNN-MP-BSBL}(b). The weighted message passing is represented by a weighting matrix $\mathbf{W}$, whose $(i,j)$-th entry is non-zero if the $i$-th input node is connected to the $j$-th output node.
\subsection{Weighted Message Passing}
\noindent\underline{\textbf{Layer 1}}: Layer 1 is the input within one iteration block.

\noindent\underline{\textbf{Layer 2}}: Layer 2 is the auxiliary layer $A^l_1$ and the output $\mathbf{O}^l_{A_1}$ is derived as follows,
\begin{equation}\label{W_OAv}
\mathbf{O}^l_{A^v_1}=\frac{\mathbf{P}^2}  {\frac{1}{\hat{\lambda}^{l-1}}\times \mathbf{W}^{l}_{\lambda\to A^v_1}+\mathbf{v}^{l-1}_{\delta \to z}\times\mathbf{W}^l_{v_\delta\to A^v_1}}
\end{equation}
\begin{equation}\label{W_OAm}
\mathbf{O}^l_{A^m_1}=\frac{\mathbf{P}^H.*\left( \mathbf{y}\times\mathbf{W}^{l}_{y\to A^m_1} \!-\!\mathbf{m}^{l-1}_{\delta\to z}\times\mathbf{W}^{l}_{{m_\delta}\to A^m_1}  \right)} {\frac{1}{\hat{\lambda}^{l-1}}\times \mathbf{W}^{l}_{\lambda\to A^m_1}\!+\!\mathbf{v}^{l-1}_{\delta \to z}\times\mathbf{W}^l_{v_\delta\to A^m_1}}
\end{equation}
The fraction and $.*$ operations are performed elementwise while the $\times$ operation is the matrix multiplication.

\noindent\underline{\textbf{Layer 3}}: The output $\mathbf{I}^l_Q$ of Layer 3 is derived as follows,
\begin{equation}\label{W_Q}
\begin{split}
\mathbf{v}^l_Q&=\frac{1}{\mathbf{O}^l_{A^v_1}\times\mathbf{W}^{l}_{A^v_1\to v_Q}}\\
\mathbf{m}^l_Q&=\mathbf{v}^l_Q.*\left( \mathbf{O}^l_{A^m_1} \times \mathbf{W}^{l}_{A^m_1 \to m_Q} \right)\!+\!\mathbf{m}^{l-1}_{\overline{h}} \!\times \! \mathbf{W}^l_{\overline{h} \to Q}
\end{split}
\end{equation}

\noindent\underline{\textbf{Layer 4}}: The output $\mathbf{I}^l_{\overline{h}}$ of Layer 4 are derived as follows,
\begin{equation}\label{W_h}
\begin{split}
\mathbf{v}^l_{\overline{h}}&\!=\!\!\frac{1}{\frac{1}{\mathbf{v}_Q^l} + \left(\mathbf{\hat{\gamma}}^{l-1}\bigotimes \mathbf{1}_{d_c}\right) \times \mathbf{W}^l_{\gamma}}\\
\mathbf{m}^l_{\overline{h}}&\!=\!\!\frac{\mathbf{m}^l_Q}{\mathbf{1}_{d_c\!K}\!\!\times\!\! \mathbf{W}_{1\to\overline{h}} \!+\! \left(\mathbf{v}^l_Q .\!*(\mathbf{\hat{\gamma}}^{l-1}\bigotimes \mathbf{1}_{d_c})\right) \!\!\times\! \!\mathbf{W}^l_{v\gamma \to \overline{h}}}
\end{split}
\end{equation}

\noindent\underline{\textbf{Layer 5}}: The output $\hat{\gamma}^l=\{\hat{\gamma}^l_k, \forall k\}$ is derived as follows,
\begin{equation}\label{W_gamma}
\mathbf{\hat{\gamma}}^l=\frac{\mathbf{a}+d_c+1}{\mathbf{b}+|\mathbf{m}^l_{\overline{h}}|^2\times \mathbf{W}^l_{m_{\overline{h}}\to\gamma} + \mathbf{v}^l_{\overline{h}} \times \mathbf{W}^l_{v_{\overline{h}}\to\gamma}}
\end{equation}

\noindent\underline{\textbf{Layer 6}}: The output $\mathbf{O}^l_{A^v_2}$ and $\mathbf{O}^l_{A^m_2}$ of Layer 6 is derived as
\begin{equation}\label{OA2}
\begin{split}
\mathbf{O}^l_{A^v_2}&=\mathbf{P}^2.*\mathbf{v}^l_{\overline{h}}\\
\mathbf{O}^l_{A^m_2}&=\mathbf{P}.*\mathbf{m}^l_{\overline{h}}
\end{split}
\end{equation}
\begin{algorithm}[t!]\setstretch{1.0}

               \caption{DNN-MP-BSBL algorithm}

               \label{alg:MP-BSBL}

               %\fontsize{8pt}{8pt}\selectfont

               \KwIn{$\mathbf{y}$, $\mathbf{\overline{P}}$, $N_{it}$, $d_c$, weighting matrices}

               %{\ \ \ \ \ \ \ \ trained  in equations (\ref{W_OAv}) to (\ref{W_lambda})}

               \KwOut{$\hat{\mathbf{h}}_{\text{DNN}}$, the index set of active user IDX}

               {\textbf{Initialize:}$\hat{\lambda}^0=10^3;\mathbf{\hat{\gamma}}^0=\mathbf{1};\mathbf{v}^0_{\delta\to z}=\mathbf{1}, \mathbf{m}^0_{\delta\to z}=\mathbf{0};\mathbf{m}^0_{\overline{h}}=\mathbf{0}$}

               %{\ \ \ \}

               %{\ \ \ \ $$.}

               \For{$l=1:N_{it}$}{

               1. Update $\mathbf{v}^l_Q$ and $\mathbf{m}^l_Q$ by (\ref{W_OAv}), (\ref{W_OAm}), and (\ref{W_Q}).

               2. Update $\mathbf{v}^l_{\overline{h}}$ and $\mathbf{m}^l_{\overline{h}}$ by (\ref{W_h}).

               3. Update $\mathbf{\hat{\gamma}}^l$ by (\ref{W_gamma}).

               4. Update $\mathbf{v}^{l}_{\delta\to z}$ and $\mathbf{m}^{l}_{\delta\to z}$ by (\ref{OA2}) and (\ref{W_delta2z}).

               5. Update $\mathbf{v}^l_z$ and $\mathbf{m}^l_z$ by (\ref{Wz}).

               6. Update the noise precision $\hat{\lambda}^{l}$ by (\ref{W_lambda}).
               }
    \textbf{return:} IDX=find$\left(\{\left({\hat{\gamma}^l_k}\right)^{-1}\}>\gamma_{th}\right)$,

 %   {\ \ \ \}

    {\ \ \ \ $\hat{\mathbf{h}}_{\text{DNN}}=\{\mathbf{0}, k \notin \text{IDX}\} \bigcup \{m^l_{\overline{h}_{k,d}}, k \in \text{IDX}, d=1,\ldots,d_c\}$.}
\end{algorithm}
\noindent\underline{\textbf{Layer 7}}: The output  $\mathbf{I}^l_{\delta\to z}$ of Layer 7 is
\begin{equation}\label{W_delta2z}
\begin{split}
\mathbf{v}^l_{\delta\to z}&\!=\!\mathbf{O}^l_{A^v_2} \times \mathbf{W}^l_{A_2^v \to v_\delta}\\
\mathbf{m}^l_{\delta\to z}&\!=\!\mathbf{O}^l_{A^m_2} \times \mathbf{W}^l_{A_2^m \to m_\delta}\\
&\!-\!\frac{\mathbf{v}_{\delta\to z}.*\left(\mathbf{y}\times \mathbf{W}^l_{y\to \delta}-\mathbf{m}^{l-1}_{\delta\to z}\times \mathbf{W}^l_{m_\delta \to m_\delta} \right)}{\left(\hat{\lambda}^{l-1}\right)^{-1}\times \mathbf{W}^l_{\lambda \to \delta}+\mathbf{v}^{l-1}_{\delta\to z}\times \mathbf{W}^{l}_{v_\delta \to m_\delta}}
\end{split}
\end{equation}

\noindent\underline{\textbf{Layer 8}}: The output $\mathbf{I}^l_{z}$ of Layer 8 is
\begin{equation}\label{Wz}
\begin{split}
\mathbf{v}^l_z&\!=\!\frac{1}{\hat{\lambda}^{l-1}\times\mathbf{W}^l_{\lambda\to z}+\frac{1}{\mathbf{v}^l_{\delta\to z}}\times\mathbf{W}_{{v_\delta}\to v_z}}\\
\mathbf{m}^l_z&\!=\!\mathbf{v}^l_z .\!*\!\left( (\mathbf{y}.*\hat{\lambda}^{l-1})\!\times\! \mathbf{W}^l_{y\lambda\to z} \!\!+\! \!\frac{\mathbf{m}^l_{\delta\to z}}{\mathbf{v}^l_{\delta\to z}}\!\times\! \mathbf{W}^l_{mv\to z}\right)
%\frac{1}{\hat{\lambda}^{l-1}\times\mathbf{W}^l_{\lambda\to z}+\frac{1}{\mathbf{v}^l_{\delta\to z}}\times\mathbf{W}_{{v_\delta}\to v_z}}
\end{split}
\end{equation}

\noindent\underline{\textbf{Layer 9}}: The output $\hat{\lambda}^l$ of Layer 9 is,
\begin{equation}\label{W_lambda}
\hat{\lambda}^l\!=\!\frac{L_tN}{\left(\mathbf{m}^l_z \!\times\! \mathbf{W}^l_{m_z\to\lambda}\!-\!\mathbf{y}\!\times\!\mathbf{W}^l_{y\to\lambda}\right)^2\!\times\! \mathbf{1}^T_{N L_t}\!+\!\mathbf{v}^l_z\!\times\! \mathbf{W}^l_{v_z\to\lambda}}
\end{equation}
\subsection{Summary of the Proposed DNN-MP-BSBL Algorithm}
Finally, the proposed DNN-MP-BSBL algorithm is summarized in Algorithm \ref{alg:MP-BSBL}. The mean square error (MSE) $\|\hat{\mathbf{h}}_\text{DNN}-\mathbf{h}\|_2^2$ is employed as the loss function for the training period, while the Normalized MSE (NMSE) $\|\hat{\mathbf{h}}_\text{DNN}-\mathbf{h}\|_2^2/\|\mathbf{h}\|_2^2$ is considered for the simulations in the testing period.
\section{Simulations}\label{simulationSec}
Parameters for the simulations are listed in Table \ref{SimulationParameters}. We consider a crowded NORA system with low-latency requirement, i.e., $N_{it}\leq20$. The NMSE performances of the LS-AMP-SBL esimator \cite{LSAMPSBL}, the BOMP estimator (with known active user number) \cite{MPBSBL} and the GA-MMSE estimator (with known user activity) are also considered. The NMSE performance of GA-MMSE estimator serves as the lower bound.
\begin{table}
\renewcommand\arraystretch{1.0}
\caption{Related Parameters for Simulations}\vspace{-0.1cm}
\centering
\begin{tabular}{ccc}
  \Xhline{1.2pt}
Parameter&Symbol&Value\\
  \hline
User number&$K$&110\\
Subcarrier number&$N$&8\\
Pilot length&$L_t$&11\\
Spreading factor&$d_c$&4\\
Activation probability for each user&$P_a$&0.1\\
UAD threshold&$\gamma_{th}$&0.1\\
Size of training set&&$10^5$\\
Size of test set&&$10^5$\\
Size of mini-batch&&200\\
Epoch number&&20\\
Learning rate&&$10^{-3}$\vspace{-0.1cm}\\
  \Xhline{1.2pt}
\end{tabular}
\label{SimulationParameters}
\end{table}

\begin{figure}
  \centering
  \includegraphics[width=0.77\linewidth]{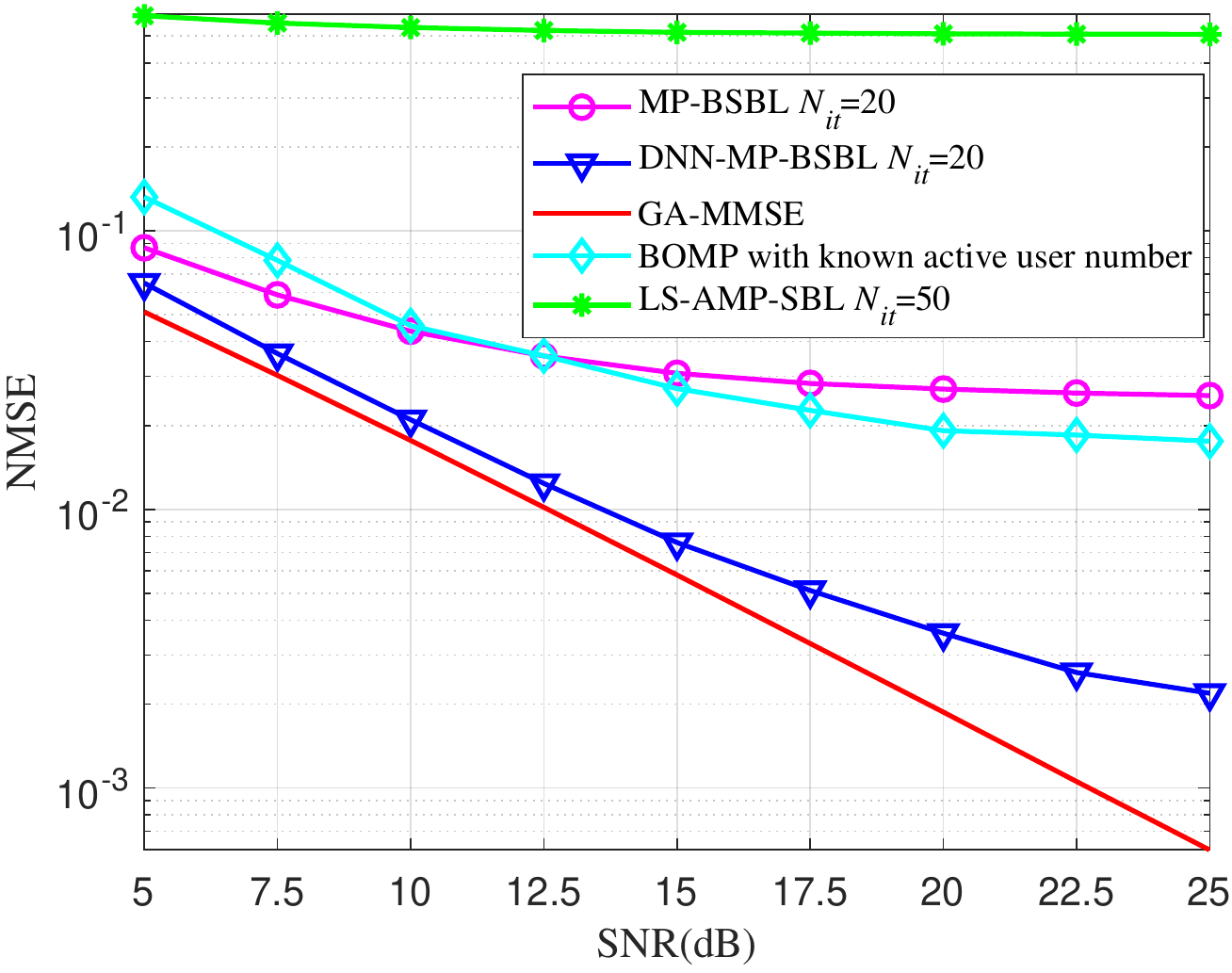}\vspace{-0.35cm}\\
  \caption{NMSE performance with different SNR.}\vspace{-0.35cm}\label{changeSNR}
\end{figure}

The simulation results are shown in Fig. \ref{changeSNR}, in crowded NORA systems, both the MP-BSBL algorithm and the BOMP estimator diverge from the NMSE lower bound as SNR increases, and the LS-AMP-SBL algorithm fails to work even with 50 iterations. By contrast, the DNN-MP-BSBL algorithm could closely approach the lower bound within a wide range of SNR. Therefore, the DNN-MP-BSBL algorithm requires fewer iterations and provides better NMSE performance, indicating its advantages in crowded NORA system with low-latency requirement.
\section{Conclusions}\label{conclusions}
A DNN-MP-BSBL algorithm was proposed in this paper for the joint UAD and CE problem in grant-free NORA systems. The iterative message passing process is transferred from a factor graph to a DNN, while weights are imposed on the messages and trained to improve the UAD and CE accuracy. Simulation results showed that the NMSE performance of the DNN-MP-BSBL algorithm could approach the lower bound in a feasible number of iterations, indicating its advantages for low-latency NORA systems.
\section*{Acknowledgement}
This work was supported by the NSFC under grant 61750110529.
\bibliographystyle{IEEEtran}

%\bibliographystyle{plain}
%\bibliography{Reference}

\end{document}